\documentstyle[12pt,psfig]{l-aa}

\def\approxgt{\mathrel{\hbox{\rlap{\lower.55ex \hbox {$\sim$}}
        \kern-.3em \raise.4ex \hbox{$>$}}}}
\def\approxlt{\mathrel{\hbox{\rlap{\lower.55ex \hbox {$\sim$}}
        \kern-.3em \raise.4ex \hbox{$<$}}}}

\def\sax{{\it BeppoSAX}}
\def\src{\mbox{Cyg\,X-2}}
\def\degmark{^\circ}

\begin{document}

\thesaurus{(02.01.2; 08.02.1; 08.09.1; 08.14.1; 13.25.3)}

\title{Low-energy line emission from Cygnus X-2 observed with the \sax\ LECS}
 
\author{E.~Kuulkers\inst{1,2} \and A.N.~Parmar\inst{1} 
\and A.~Owens\inst{1} \and T.~Oosterbroek\inst{1} \and U.~Lammers\inst{1}}

\institute{Astrophysics Division, Space Science Department of ESA, 
ESTEC, P.O. Box 299, 2200 AG Noordwijk, The Netherlands
\and Astrophysics, University of Oxford, Nuclear and Astrophysics Laboratory,
Keble Road, Oxford, OX1 3RH, UK}

\date{Received ; accepted}
\offprints{E.~Kuulkers: erik@astro.ox.ac.uk}
\maketitle

\begin{abstract}

We present a 0.2--10\,keV spectrum of the
low-mass X-ray binary Cygnus~X-2 obtained using the Low Energy
Concentrator Spectrometer on-board \sax.
The spectrum can be described by a cut-off power-law 
model with absorption of (2.28$\pm$0.07)$\times$10$^{21}$\,atoms\,cm$^{-2}$, 
a power-law index of 0.78$\pm$0.02 and a cut-off energy of 4.30$\pm$0.08\,keV
(68\% confidence errors), except at energies near 
$\sim$1\,keV where excess emission is present.
This can be modeled by a broad Gaussian line feature with an
energy of 1.02$\pm$0.04\,keV, a full width half-maximum of
0.47$\pm$0.07\,keV and an equivalent width of 74$\pm ^{25}_{11}$\,eV.
This result confirms earlier reports of line emission near 1\,keV
and shows the intensity and structure of the feature to be variable.

\end{abstract}

\keywords{accretion $-$ binaries:close $-$ stars:individual (\src) 
$-$ stars:neutron $-$ X-rays: general} 

\section{Introduction}

Cygnus~X-2 is a bright persistent low-mass X-ray 
binary (LMXRB), whose X-ray spectrum has been studied 
from $\sim$0.1 to several hundred keV. Together with some of the other 
bright persistent LMXRB sources it is classified as 
a ``{\it Z\/}'' source 
(Hasinger \&\ van der Klis \cite{hk89}). The changes in the X-ray spectral
shape of {\it Z\/} sources are subtle, but in an X-ray color-color diagram
the sources trace out {\it Z\/}-like shaped patterns. They move through the 
{\it Z\/} in a smooth manner without jumping from branch to 
branch.
{\it Z\/} sources are thought to be accreting material at near-Eddington 
rates via an 
accretion disk onto a neutron star (e.g. Hasinger et al.\ 1990).

The Italian-Dutch satellite
\sax\ is the first mission to simultaneously observe in the 
0.1--300\,keV energy
range using a complementary payload of instruments (Boella et al.\ 1997).
The Low-Energy Concentrator Spectrometer (LECS)
is sensitive in the energy range \mbox{0.1--10}\,keV (Parmar et al.\ 1997). 
Its unique design utilizes a driftless gas scintillation 
proportional counter to make the lowest 
energies accessible with a good energy resolution while providing
16\,$\mu$s time resolution and moderate spatial resolution. 
The LECS has a circular field of view of 37$'$ 
diameter and a 0.1--10\,keV background counting rate of
9.7$\times$10$^{-5}$\,arcmin$^{-2}$\,s$^{-1}$.
The LECS energy resolution, $\Delta$E/E, is 19\% at 1~keV and
varies as E$^{-0.42}$.
In this {\it Letter} results from the LECS are 
presented for one of the Science Verification Phase targets, \src. 
We find that the 0.2--10\,keV spectrum can be described by an absorbed 
cut-off power-law model, with additional emission at energies 
near $\sim$1\,keV.

\begin{table*}
\caption[]{Fit results to the LECS \src\ spectrum$^a$}
\begin{tabular}{llll}
\noalign {\smallskip}
\hline
\noalign {\smallskip}
\multicolumn{1}{c}{~} & \multicolumn{2}{c}{Model} & \multicolumn{1}{c}{~} \\
& Cut-off power-law & Cut-off power-law & Units \\
      &                   & + Gaussian        &       \\
\noalign{\smallskip}
\hline
\noalign{\smallskip}
N$_{\rm H}$     & $2.22 \pm 0.04$   & $2.28 \pm 0.07$ & 10$^{21}$\,H\,atoms\,cm$^{-2}$ \\
$\gamma$        & $0.876 \pm 0.012$ & $0.774 \pm 0.019$ &     \\
E$_{\rm cut}$   & $4.77 \pm 0.09$   & $4.30 \pm 0.08$ & keV \\
K               & $1.267 \pm 0.008$ & $1.220 \pm 0.011$ & photons\,cm$^{-2}$\,s$^{-1}$\,keV$^{-1}$ at 1\,keV \\
E$_{\rm c}$     & & $1.02 \pm 0.04$ & keV \\
FWHM            & & $0.47 \pm 0.07$ & keV \\
EW              & & $74 \pm ^{25}_{11}$ & eV \\
$\chi^2_{\nu}$/dof & 1.45/925 & 1.20/922 &   \\
\noalign{\smallskip}
\hline
\multicolumn{4}{l}{\footnotesize $^a$\,Uncertainties are given at the 68\%\ confidence level.}
\end{tabular}
\end{table*}

\section{Observation and Results}

\sax\ observed \src\ between 1996 July 23 00:19 and 18:18~UTC. 
Good data were selected from intervals when the minimum elevation angle
above the Earth's limb was $>$4$\degmark$ and when the instrument
configuration was nominal using the SAXLEDAS 1.4.0 data analysis package
(Lammers \cite{l97}).  
The LECS was only operated during satellite night-time giving a 
total on-source exposure of 7.1\,ks.
A spectrum was extracted centered on 
the mean source position 
using the standard LECS extraction radius of 8$'$ and 
the appropriate response matrix was generated.
Background subtraction was performed using a standard blank field 46\,ks 
exposure, but is not critical for such a bright source.
The spectrum was rebinned to have at least 20 counts per channel. Channels
below 0.2\,keV were discarded since the source is absorbed below this energy.
The mean LECS count rate observed from \src\ is 
67\,s$^{-1}$, which corresponds to an observed 
1--10\,keV flux of 7.4$\times$10$^{-9}$\,erg\,cm$^{-2}$\,s$^{-1}$.

The ground calibration of the LECS is discussed in Parmar et al. (1997).
The effective area $\approxlt$2~keV is primarily limited by the 
entrance window transmission, and at higher energies by the
loss of reflectivity of the mirror system. Great care was taken to determine
the entrance window transmission 
using monochromatic X-ray sources (Bavdaz et al. 1994). 
The LECS spectral response 
was updated following the first \sax\ observation of the Crab
Nebula on 1996 September 6. When the 
LECS spectrum is fit by an absorbed power-law model, the 
individual energy channel residuals are $\approxlt$5\%. 
The calibration status of the \sax\ instruments, derived from the 
above observation, is discussed in Cusumano et al. (1997).  

The \src\ spectrum can be modeled by a cut-off 
power-law (${\rm K E^{-\gamma} \exp(E_{cut}/kT}$)) together with low energy
absorption, N$_{\rm H}$, which gives a $\chi ^2_{\nu}$ of 1.45 for 925 degrees
of freedom (dof), see also Table~1. Examination of the residuals shows 
excess counts near 1\,keV which are unlikely to originate from calibration 
uncertainties or other instrumental effects. 
These excess counts can be modeled by a blackbody of temperature
0.181$\pm0.007$~keV giving a $\chi ^2_{\nu}$ of 1.25 for 923 dof. 
However, previous observations of \src\ have required much
higher blackbody temperatures of 1.0--1.5~keV (e.g.\ Hasinger et al.\ 1990),
and it is therefore unlikely that this feature is the blackbody previously
observed from the source.
Replacing the blackbody with a broad Gaussian line feature
with an energy, E$_{\rm c}$, of $1.02 \pm 0.04$~keV, a
full width at half-maximum, FWHM, of 0.47$\pm$0.07\,keV
and an equivalent width, EW, of (74$\pm ^{25}_{11}$)\,eV in the fit 
gives a better fit with a $\chi ^2_{\nu}$ of 1.20 for 922 dof.
An {\it F\/}-test indicates that this difference is significant, but
only at the 1$\sigma$ level. 
Higher energy resolution \src\ observations than that reported here 
(e.g.\ using the {\it Einstein} 
Objective Grating Spectrometer (OGS); see Vrtilek et al.\ 1988),
require the presence of multiple line features near 1~keV.
This supports the interpretation of the excess emission seen in the 
LECS as unresolved line emission.
Table~1 gives the best-fit parameters for the Gaussian line model. Fig.~1
shows the observed count spectrum and the best-fit model when the 
normalization of the emission feature is set to zero.

In addition to the feature at $\sim$1\,keV, there is evidence for
weak narrow-line features at 1.54$\pm$0.03, 2.01$\pm$0.02, 
and 2.61$\pm$0.02\,keV in the LECS spectrum (see Fig.~1). 
Including all three lines in the fit reduces the $\chi^2_{\nu}$
to 1.15 for 916 dof.
The energies of two of these lines
are consistent with Ly$\alpha$ transitions of H-like Si~{\sc xiv} and
S~{\sc xvi} at 2.00 and 2.62\,keV, respectively (see e.g.\ Raymond 1993).
However, we caution that these features have
energies close to those of the mirror Au
edges, where the LECS calibration is more uncertain.   
There is no evidence for Gaussian line emission near
6.7\,keV, with a 3$\sigma$ upper limit on the equivalent width of 62\,eV
for a narrow line.

\begin{figure*}
 \centerline{\hbox{
  \psfig{figure=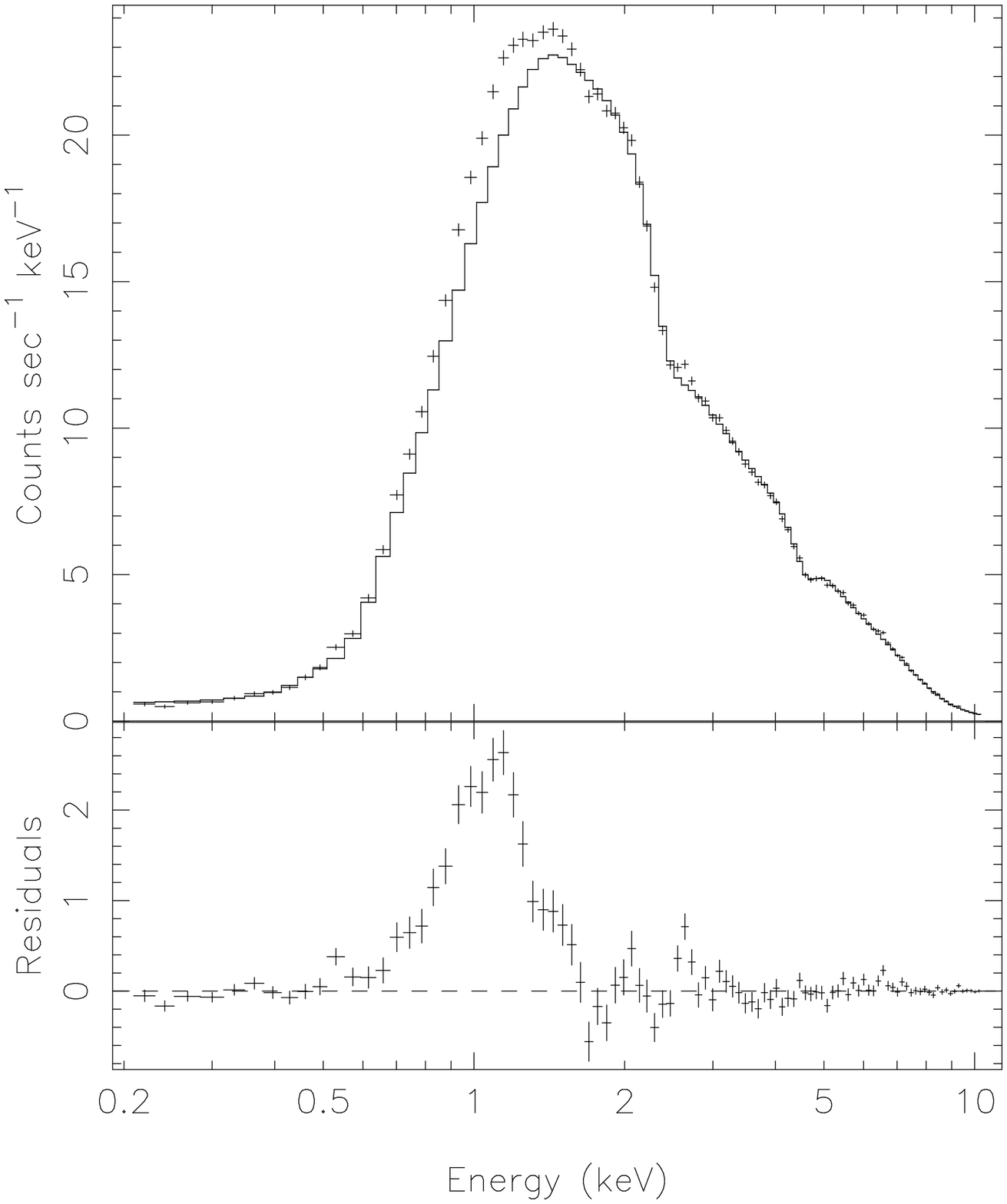,bbllx=24pt,bblly=47pt,bburx=534pt,bbury=656pt,height=9.5cm}
   \hspace{0.5cm}
  \psfig{figure=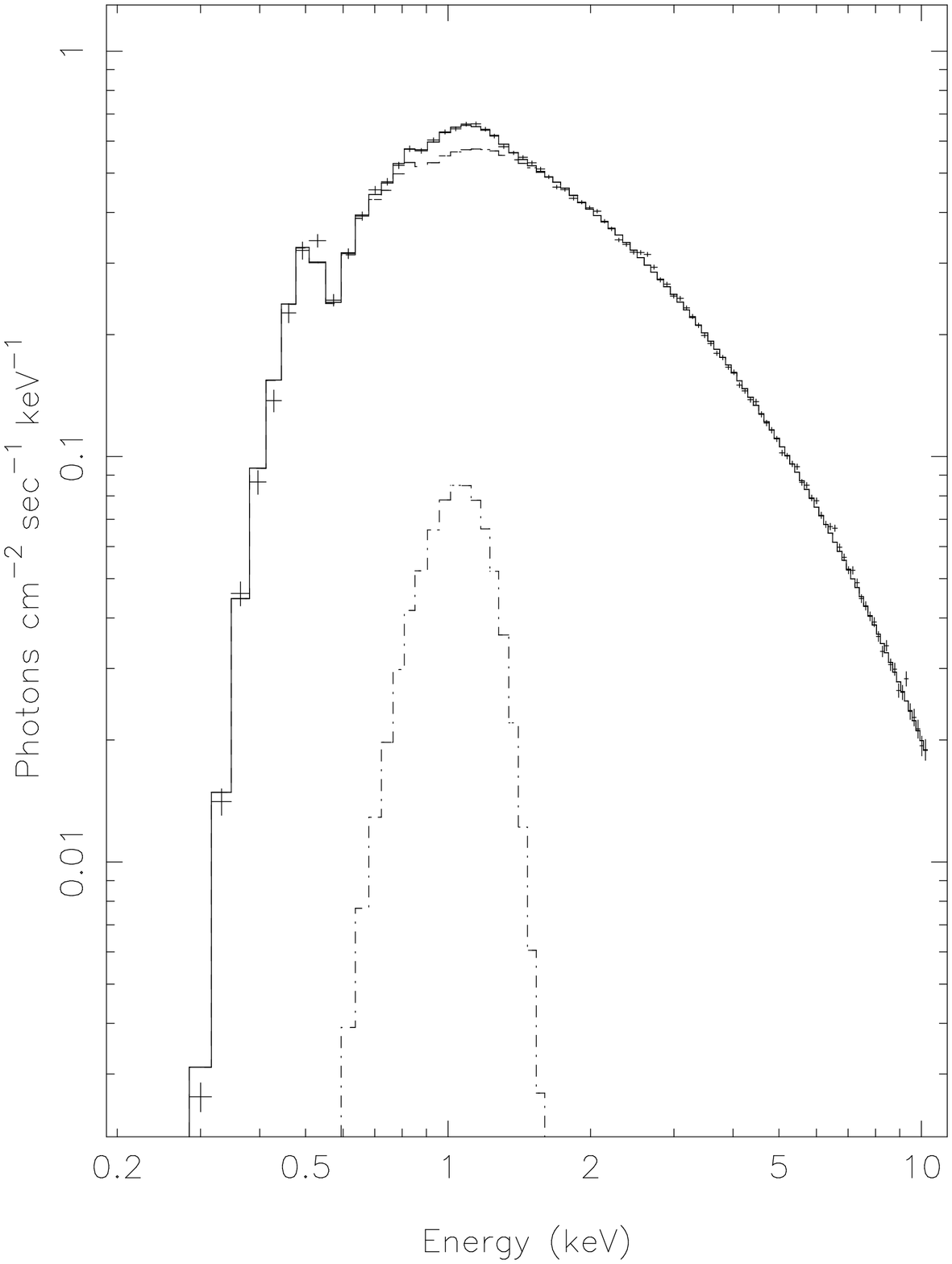,bbllx=24pt,bblly=25pt,bburx=534pt,bbury=702pt,height=9.5cm}
 }}
\caption[]{The observed LECS spectrum together with the best-fit cut-off 
power-law model with the normalization of the 1.02\,keV emission feature
set to zero (upper left panel). The lower left panel shows the residuals.
The right panel shows the inferred photon spectrum and the model prediction
including the 1.02\,keV feature} 
\end{figure*}

Vrtilek et al.\ (1988) have compiled the results of {\it Einstein} 
OGS and medium-resolution Solid State Spectrometer 
(SSS) observations of \src. The OGS data indicate the presence of four narrow
emission lines between 0.74 and 1.12\,keV (see Table~2). If the lower 
resolution SSS data is fit with the same set of line energies,  
the results are broadly consistent with the OGS results 
except that the EW of each of the
lines is a factor $\sim$2 larger (Vrtilek et al.\ 1988).
In order to investigate whether the excess emission seen by the LECS
is consistent with the same blend of lines as seen by {\it Einstein},
the broad Gaussian emission feature was replaced by four narrow 
lines at the energies given in Table~2. The resulting $\chi^2_{\nu}$ is 1.24
for 921 dof. There is no evidence for the presence of the 0.74\,keV
Fe~{\sc xvii} and 0.77\,keV O~{\sc viii}/Fe~{\sc xviii} features with 
95\%\ confidence
upper limits on the EW of 14 and 16\,eV, respectively. The EW of the
0.96\,keV Fe~{\sc xx}/Ni~{\sc xx} feature of 14$\pm$3\,eV is similar
to the values seen by the OGS and SSS of 12$\pm$5 and 
30$\pm$10\,eV, respectively. However, the EW of the 1.12\,keV
line of 24$\pm$3\,eV is larger than measured by both the
OGS (5$\pm$2\,eV) and SSS (10$\pm$3\,eV).
We caution that the narrow-line fit accounts for only about half
the line flux observed, suggesting the presence of other emission
features in this energy range.
 
\begin{table}
\caption[]{A comparison of LECS and {\it Einstein} line energies and equivalent widths$^a$}
\begin{tabular}{llccc}
\noalign {\smallskip}
\hline
\noalign {\smallskip}
Line   & Line           & \multicolumn{3}{c}{EW} \\
energy & identification &  OGS & SSS & LECS \\
(keV)  &                &  (eV) & (eV) & (eV) \\
\noalign{\smallskip}
\hline
\noalign{\smallskip}
0.74 & Fe~{\sc xvii}               & $4 \pm 2$ & $10 \pm 3$ 
& $<$14\\
0.77 & O~{\sc viii}/Fe~{\sc xviii} & $6 \pm 3$ & $10 \pm 3$ 
& $<$16 \\
0.96 & Fe~{\sc xx}/Ni~{\sc xx}     & $12 \pm 5$ & $30 \pm 10$ 
& $14 \pm 3$ \\
1.12 & Fe~{\sc xvii}/ & $5 \pm 2$ & $10 \pm 3$ 
& $24 \pm 3$ \\
     &              Fe~{\sc xxiii}--{\sc xxiv} & \\
\noalign{\smallskip}
\hline
\multicolumn{5}{l}{\footnotesize $^a$\,The OGS and SSS results are taken 
from Vrtilek et al.}\\
\multicolumn{5}{l}{\footnotesize (1988) as are the line energies and 
identifications.}\\
\end{tabular}
\end{table}

\section{Discussion}

Models for the continua of many bright LMXRB
sources often require at least two components, e.g. a blackbody 
and a cut-off power-law model 
(White et al. 1988). A blackbody component is not required
to fit the LECS spectrum. The absence of the blackbody component
when \src\ is on the upper left part of its horizontal branch of the {\it Z\/}
has been noted by Kuulkers et al.\ (\cite{kkp95}).
The blackbody component appears to be present 
on all other parts of the {\it Z\/} (Hasinger et al.\ \cite{hke90};
Kuulkers et al.\ \cite{kkp95}). 

Although the timing properties can
in principle be used to determine the position of \src\ in its 
{\it Z\/} variability pattern, there are insufficient
counts in the LECS data to uniquely determine it.
The intensity of \src\ is normally a factor of $\sim$2 higher 
(e.g. Hasinger et al.\ 1990; Kuulkers et al.\ 1995)
than during the \sax\ observation. Such low intensities are only
reached during the so-called ``low intensity'' states of \src\
(\mbox{Kuulkers} et al.\ 1996).
Inspection of the {\it Rossi} X-ray Timing Explorer All 
Sky Monitor (ASM) light curves (Wijnands et al.\ 1996)
indicates that the \sax\ observation occured 
5--10 days before \src\ entered a probable low intensity state,
suggesting that the source was transitioning from a 
medium to a low intensity state. We note that a similar transition may have 
been seen by Vrtilek et al.\ (1986; their ``medium state B''). 
Vrtilek et al.\ (1986) fit the 1--20\,keV {\it Einstein} Monitor Proportional 
Counter (MPC) spectra obtained during this state with a thermal
bremsstrahlung model of kT $\sim$7.0--8.7\,keV and 
N$_{\rm H}$ $\sim$2.3--3.5$\times10^{21}$\,H\,atoms\,cm$^{-2}$.
A comparison with the best-fit parameters obtained when the same model
is fit to the 1--10\,keV LECS spectrum of kT$\sim$9.2\,keV and 
${\rm N_H}$$\sim$3.5$\times 10^{21}$\,cm$^{-2}$
indicates that the spectral shape was similar on both occasions.

The presence of excess emission near 1.0\,keV was first
suggested following a rocket flight in 1971 (Bleeker et al.\ \cite{bdy72}).
As well as the OGS and SSS results reported in Vrtilek et al.\ 
(1986, 1988; see also Kallman et al.\ \cite{kvk89} and 
Smale et al.\ \cite{saw94}), Branduardi-Raymont et al.\ (\cite{bce84}) using
the {\it Ariel~V} Experiment~C, Chiappetti et al.\ (\cite{ctb90}) using the 
EXOSAT Channel Multiplier Array, Lum et al.\ (\cite{lcc92}) using the 
{\it Einstein} Focal Plane Crystal Spectrometer, 
Smale et al.\ (\cite{sdm93}) using the Broad Band X-ray Telescope and
Smale et al.\ (\cite{saw94}) using the {\it ASCA} Solid State Imaging 
Spectrometer, all report evidence for excess emission near 1\,keV. 
These reports indicate that the strength of the 
excess emission varies from observation to observation.
No excess emission near 1\,keV was reported by Pravdo (\cite{p83}),
Hirano et al.\ (\cite{hhk84}), and Predehl \&\ Schmitt (\cite{ps95}) 
using lower energy resolution data obtained with the HEAO\,1\,A2, 
{\it Hakucho} proportional
counter, and {\it ROSAT} Position Sensitive Proportional Counter instruments, 
respectively.

Line emission at energies near 1\,keV is not unique to \src. Similar features
have been observed in other LMXRB, e.g.\
in the {\it Z\/} source \mbox{Sco\,X-1} (see Vrtilek et al.\
\cite{vms91}), and in the accreting pulsars \mbox{Her\,X-1} 
(McCray et al.\ 1982; Oosterbroek et al.\ 1997) and \mbox{4U\,1626$-$67} 
(Angelini et al.\ 1995; Owens et al.\ 1997).
The feature is most probably due to a combination of unresolved 
Fe~L-shell and Ne~K-shell line emission (Chiappetti et al.\ \cite{ctb90}; 
Vrtilek et al.\ \cite{vms91}; Lum et al.\ \cite{lcc92}; 
Smale et al.\ \cite{sdm93}, \cite{saw94}; Angelini et al.\ 1995).
Vrtilek et al.\ (1988) modeled the $\sim$1\,keV excess emission seen
from \src\ with {\it Einstein} using four 
narrow emission lines which they identify with Fe~{\sc xvii}, 
O~{\sc viii}/Fe~{\sc xvii}, Fe~{\sc xx}/Ni~{\sc xx} and 
Fe~{\sc xvii}/Fe~{\sc xxii}--{\sc xxiv} features (see also Kallman et al.\ 
\cite{kvk89}). 
The LECS spectrum was fit with the same model,
and we find that the Fe~{\sc xvii} and O~{\sc viii}/Fe~{\sc xvii} 
features are not present
(although the 95\% confidence upper limits are consistent
with the {\it Einstein} detections), the EW of the Fe~{\sc xx}/Ni~{\sc xx} 
feature is similar to the values reported in Vrtilek et al.\ (1988), 
and the EW of the Fe~{\sc xvii}/Fe~{\sc xxii}--{\sc xxiv} feature is 
significantly greater (see Table~2).
These differences imply that the shape of the feature, as well as its
overall EW, varies from observation to observation.
It is notable that the narrow line model only accounts for about 
half the line flux of the excess emission as seen with the LECS, implying the 
presence of other features near 1\,keV.

The Fe-L emission may be produced by photoionization of the surface of the
accretion disk and the accretion disk corona
by the strong X-ray continuum flux emanating from the central regions 
(see e.g.\ Kallman et al.\ \cite{kvk89}; Raymond \cite{r93}).
Recent calculations by Kallman (\cite{k95}) show that the ratio of the 
equivalent widths of the {\mbox Fe-L} and Fe-K lines should be close to unity 
for \src. Our measurements indicate that this ratio is 
$\approxlt$1.2 (3$\sigma$), consistent with these calculations.

The strength of the low-energy feature is expected to be
dependent on the temperature and 
density of the illuminated corona (e.g.\ Lidahl \cite{lko90}; 
Kallman \cite{k95}), and probably indirectly on the mass 
accretion rate onto the neutron star which probably influences the X-ray 
continuum and 
shape of the inner accretion disk. Since these parameters can vary from 
observation to observation,  it is not surprising that 
the strength and shape of the excess emission near 1\,keV is variable.

\begin{acknowledgements}
We thank R.C.~Butler, L.~Piro and the staff of the \sax\ Science Data Center.
The \sax\ satellite is a joint Italian and Dutch programme. 
T. Oosterbroek acknowledges an ESA Fellowship.
\end{acknowledgements}

\end{document}